\documentclass[ArXiv,AXISWP,accept,moreauthors,pdftex,10pt,letterpaper]{Definitions/axis} 

\firstpage{1} 
\pubvolume{xx}
\issuenum{1}
\articlenumber{5}
\pubyear{2023}
\copyrightyear{2023}

\Title{X-ray Redshifts for Obscured Active Galactic Nuclei with AXIS Deep and Intermediate Surveys }

\Author{Alessandro Peca $^{1,\dagger}$, Nico Cappelluti $^{1}$, Stefano Marchesi$^{2,3,4}$, Edmund Hodges-Kluck$^{5}$, and Adi Foord$^{6}$}

\AuthorNames{\orcidA{}Alessandro Peca, Nico Cappelluti, Stefano Marchesi,  Edmund Hodges-Kluck, Adi Foord}

\address{%
$^{1}$ \quad Department of Physics, University of Miami, Coral Gables, FL 33124, USA\\
$^{2}$ \quad Dipartimento di Fisica e Astronomia (DIFA), Università di Bologna, via Gobetti 93/2, I-40129 Bologna, Italy\\
$^{3}$ \quad Department of Physics and Astronomy, Clemson University,  Kinard Lab of Physics, Clemson, SC 29634, USA\\
$^{4}$ \quad INAF - Osservatorio di Astrofisica e Scienza dello Spazio di Bologna, Via Piero Gobetti, 93/3, 40129, Bologna, Italy\\
$^{5}$ \quad NASA/GSFC, Code 662, Greenbelt, MD 20771, USA\\
$^{6}$ \quad Department of Physics, University of Maryland Baltimore County, 1000 Hilltop Cir, Baltimore, MD 21250, USA
}

\firstnote{alessandro.peca@miami.edu} 

\abstract{This study presents the capabilities of the AXIS telescope in estimating redshifts from X-ray spectra alone (X-ray redshifts, XZs). Through extensive simulations, we establish that AXIS observations enable reliable XZ estimates for more than 5500 obscured Active Galactic Nuclei (AGN) up to redshift {\texorpdfstring{$z\sim 6$}{}} in the proposed deep (7 Ms) and intermediate (375 ks) surveys. Notably, at least 1600 of them are expected to be in the Compton-Thick regime ({\texorpdfstring{$\log N_H/\mathrm{cm^{-2}}\geq 24$}{}}), underscoring the pivotal role of AXIS in sample these elusive objects that continue to be poorly understood.
XZs provide an efficient alternative for optical/infrared faint sources, overcoming the need for time-consuming spectroscopy, potential limitations of photometric redshifts, and potential issues related to multi-band counterpart association. 
This approach will significantly enhance the accuracy of constraints on the X-ray luminosity function and obscured AGN fractions up to high redshift. {\texorpdfstring{\\}{}}
\emph{This White Paper is part of a series commissioned for the AXIS Probe Concept Mission; additional AXIS White Papers can be found at the  \href{http://axis.astro.umd.edu/}{AXIS website} with a mission overview \href{https://arxiv.org/abs/2311.00780}{here}}.
}

\begin{document}


\def\aj{AJ}%
\def\actaa{Acta Astron.}%
\def\araa{ARA\&A}%
\def\apj{ApJ}%
\def\apjl{ApJ}%
\def\apjs{ApJS}%
\def\ao{Appl.~Opt.}%
\def\apss{Ap\&SS}%
\def\aap{A\&A}%
\def\aapr{A\&A~Rev.}%
\def\aaps{A\&AS}%
\def\azh{AZh}%
\def\baas{BAAS}%
\def\bac{Bull. astr. Inst. Czechosl.}%
\def\caa{Chinese Astron. Astrophys.}%
\def\cjaa{Chinese J. Astron. Astrophys.}%
\def\icarus{Icarus}%
\def\jcap{J. Cosmology Astropart. Phys.}%
\def\jrasc{JRASC}%
\def\mnras{MNRAS}%
\def\memras{MmRAS}%
\def\na{New A}%
\def\nar{New A Rev.}%
\def\pasa{PASA}%
\def\pra{Phys.~Rev.~A}%
\def\prb{Phys.~Rev.~B}%
\def\prc{Phys.~Rev.~C}%
\def\prd{Phys.~Rev.~D}%
\def\pre{Phys.~Rev.~E}%
\def\prl{Phys.~Rev.~Lett.}%
\def\pasp{PASP}%
\def\pasj{PASJ}%
\def\qjras{QJRAS}%
\def\rmxaa{Rev. Mexicana Astron. Astrofis.}%
\def\skytel{S\&T}%
\def\solphys{Sol.~Phys.}%
\def\sovast{Soviet~Ast.}%
\def\ssr{Space~Sci.~Rev.}%
\def\zap{ZAp}%
\def\nat{Nature}%
\def\iaucirc{IAU~Circ.}%
\def\aplett{Astrophys.~Lett.}%
\def\apspr{Astrophys.~Space~Phys.~Res.}%
\def\bain{Bull.~Astron.~Inst.~Netherlands}%
\def\fcp{Fund.~Cosmic~Phys.}%
\def\gca{Geochim.~Cosmochim.~Acta}%
\def\grl{Geophys.~Res.~Lett.}%
\def\jcp{J.~Chem.~Phys.}%
\def\jgr{J.~Geophys.~Res.}%
\def\jqsrt{J.~Quant.~Spec.~Radiat.~Transf.}%
\def\memsai{Mem.~Soc.~Astron.~Italiana}%
\def\nphysa{Nucl.~Phys.~A}%
\def\physrep{Phys.~Rep.}%
\def\physscr{Phys.~Scr}%
\def\planss{Planet.~Space~Sci.}%
\def\procspie{Proc.~SPIE}%
\let\astap=\aap
\let\apjlett=\apjl
\let\apjsupp=\apjs
\let\applopt=\ao
\tableofcontents
\listoffigures

\section{Introduction}
Redshifts play a crucial role in the study of the properties and evolution of active galactic nuclei (AGN). To determine the redshift of an AGN, most solid measurements can be obtained through the spectroscopic identification of optical/near-infrared (ONIR) emission lines. However, spectroscopic redshift (spec-z) quality relies on good signal-to-noise spectra, which can be resource-intensive in terms of exposure times, especially for faint and distant AGN. 
An alternative to spec-zs is to compute photometric redshifts (photo-zs) through spectral energy distribution (SED) fitting. This is a more easily accessible method, which relies on multi-wavelength observations by comparing object fluxes in different filters with synthetic models. However, accurate photometric redshifts may be challenging to obtain for AGN because of the complex nature of their SEDs and the limited availability of filters needed to obtain reliable estimates. This challenge is further compounded by the potential dilution of the AGN emission by concurrent galaxy radiation. \citep[e.g.,][]{salvato09,salvato19}.

In the last few years, a relatively new technique based on X-ray spectra has been extensively explored to determine the redshift of AGN. These redshift estimates (X-ray redshifts, XZs) were successfully tested for obscured AGN \citep{simmonds18,peca21,sicilian22}, which may be extremely faint at ONIR wavelengths due to gas and dust along the line of sight.
On the contrary, X-ray photons can more easily escape heavy column densities ($\log N_H/{\mathrm{cm^{-2}}}>23$, \citep[e.g.,][]{buchner15, signorini23}), providing a valuable alternative for AGN that are challenging to observe spectroscopically and photometrically. Therefore, XZs offer a complementary approach that does not require ONIR datasets and avoids the complexities associated with multi-band counterpart associations \citep[e.g.,][]{salvato18}.
However, as with other redshift methods, even XZs rely on the quality of the data. Current X-ray observatories such as \textit{Chandra} and \textit{XMM-Newton} provide excellent spectra on-axis (i.e., in the center of their field of view), but as we move off-axis the PSF broadens and distorts very quickly, reducing drastically the quality of the spectra, and therefore negatively affecting the performance of the XZ method. 
In contrast, with a stable angular resolution across its entire field of view, the AXIS telescope can capture a large amount of high-quality spectra within a single pointing, more than ever before. This, coupled with its large effective area \citep{axis1,axis2,2023_AXIS_Overview}, makes it the ideal instrument for measuring XZs.
In this white paper, we explore the AXIS capabilities in deriving XZs of obscured AGN in the planned deep (7 Ms) and intermediate (375 ks) surveys, by focusing on the low-photon statistics regime to investigate the limits of this technique.

\section{Results}\label{sec:2}
We show the main results of our simulations in this section, while the detailed procedures and methods are described in Section \ref{sec:4}. 
We applied the method described in \citep{peca21} to determine the feasibility of XZs with AXIS. This method provides XZ's success rate maps as a function of redshifts, number of counts, and absorption column density ($N_H$). The success rate, or match percentage (MP), is defined as:
\begin{equation}
    \mathrm{MP
    } ({\mathrm cts}, z, N_{\rm H}) = \frac{N(z \pm \Delta z)}{N_{\rm sim}}
\end{equation}
This quantity represents the percentage of simulated sources for which the spectral fitting successfully recovered a redshift that agrees, within the errors, with the simulated value (as well as for $N_H$ and power-law normalizations). In other words, MP shows how well XZs can be estimated as a function of redshift, counts, and $N_H$, by using X-ray spectra alone. It is worth mentioning that MP is also a function of the photon index $\Gamma$, but for commonly accepted values ($\Gamma = 1.9 \pm 0.2$, \citep[e.g.,][]{nandra94,piconcelli05}) the effect on MP is negligible \citep{peca21}.
As a conservative approach \cite{simmonds18, peca21, sicilian22}, we rejected estimates where the derived XZ was consistent with the simulated one, within the errors, but where $|\Delta z| > 0.15(1 + z_{sim})$. This also applies when XZ estimates were upper or lower limits.

\begin{figure}[H] 
\centering
\includegraphics[width=16cm]{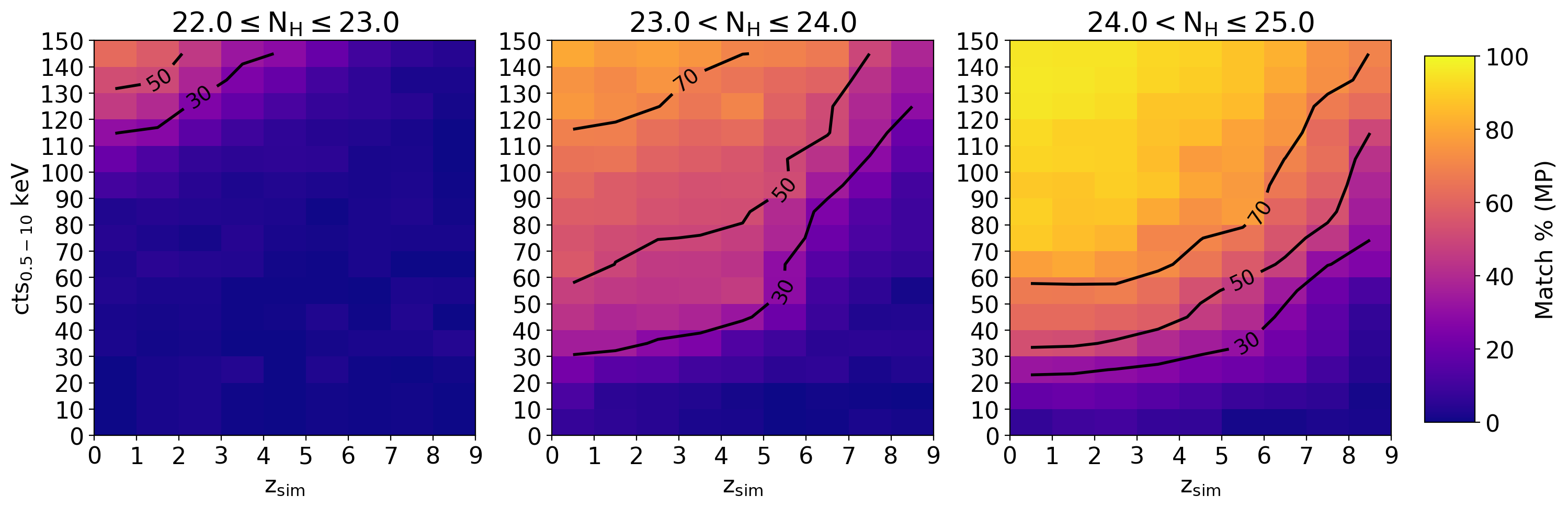}
\caption[Match percentage maps as a function of redshift and number of counts.]{Match percentage (MP) maps as a function of redshift and number of counts in the 0.5-10 keV band, for three $\log N_H$ bins. In this simulation, we used a double power-law model with the 6.4 keV Fe K$\alpha$ emission line and an exposure time of 7 Ms (see details in Section \ref{sec:4}). The solid black contours represent MP = 30, 50, and 70\%, respectively.}
\label{fig:results}
\end{figure}

In Figure \ref{fig:results} we present the results obtained for a double power-law model with $\Gamma=1.9$, $\log N_H/{\mathrm{cm^{-2}}}$ in the range 22-25, and where we included the 6.4 keV Fe K$\alpha$ emission line (see details in Section \ref{sec:4}). From the three maps, it is clear how increasing levels of absorption can provide more accurate XZ results. The reason is that the XZ estimates are driven by the 6.4 keV Fe K$\alpha$ emission line and the absorption features (7.1 keV Fe K$\alpha$ absorption edge and photoelectric absorption), which become more prominent with increasing absorption and therefore more easily identifiable by the spectral fitting \citep[e.g.,][]{iwasawa12, simmonds18}. In addition to highlighting the prominence of the main features in the X-ray spectrum when obscuration increases, these simulations provide valuable insight into the probability of accurately determining the redshift of these AGN by analyzing the X-ray spectra alone.
We chose a threshold of MP$\geq 50\%$ as the likelihood of obtaining a reliable XZ estimate. \citep{peca21,peca23} showed that this threshold is a fair compromise between having a large enough sample and spectral fit accuracy.
In particular, our results show that it is possible to compute reliable XZs up to $z\sim9$ for $\log N_H/{\mathrm{cm^{-2}}}=25$ and number of counts down to $\sim20$.
These numbers are expected to change when other AGN models are used \citep{peca21,peca23}. However, as assumed in other X-ray surveys \citep[e.g.,][]{lanzuisi15,marchesi16b}, our chosen model is a good representation of the overall shape of obscured AGN. For completeness, we discuss the effects of assuming other models in Appendix \ref{app:models}.

\section{Discussion}
\begin{figure}[H] 
\centering
\includegraphics[width=16cm]{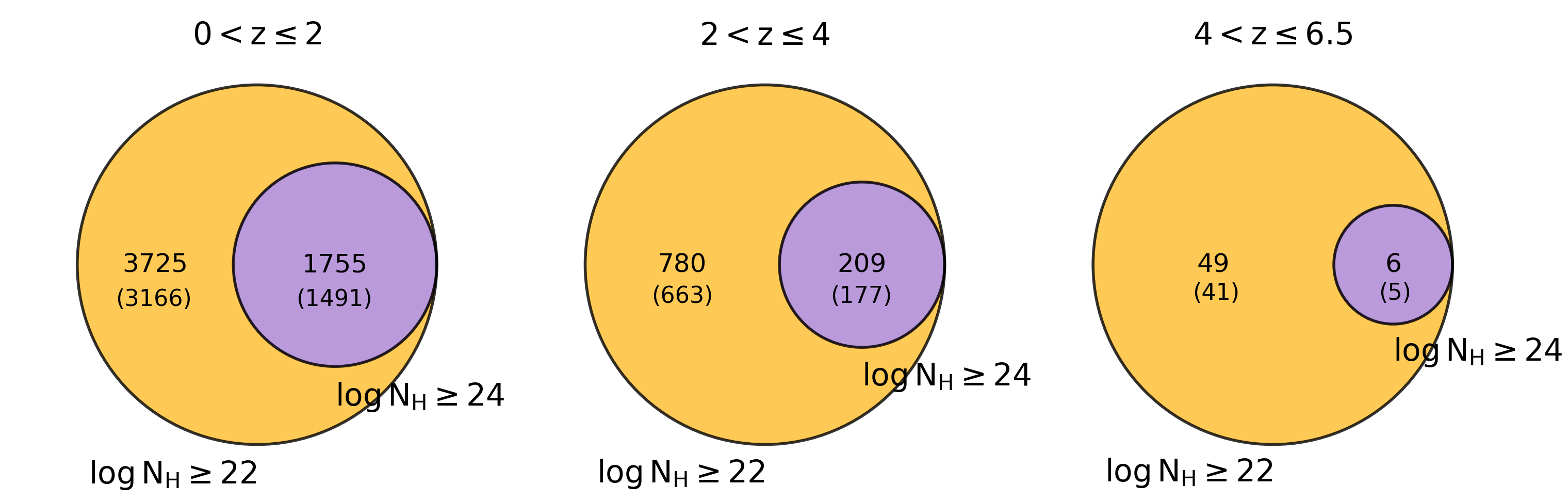}
\caption[Predicted number of XZs in the combined deep and intermediate AXIS surveys.]{Predicted number of reliable XZs in the combined deep (7 Ms) and intermediate (375 ks) planned AXIS surveys \citep{2023_AXIS_Overview}. The three Venn diagrams show different redshift bins, while the orange and purple circles show the expected number of XZs for $\log N_H/\mathrm{cm^{-2}} \geq$ 22 and 24, respectively. The provided numbers represent our projections based on a model with the 6.4 keV K$\alpha$ emission line, while the values within the parentheses indicate the scenario where no emission line is detected.}
\label{fig:results2}
\end{figure}

AXIS is expected to provide valuable X-ray redshift measurements, allowing in-depth studies of X-ray obscured AGN populations, especially for those in the Compton-thick ($\log N_H/\mathrm{cm^{-2}} \geq 24$) regime. By using the most up-to-date simulations for the deep (7 Ms) and intermediate (375 ks) AXIS surveys \cite{marchesi20,2023_AXIS_Overview}, we can predict how many reliable XZs we will be able to estimate. Figure \ref{fig:results2} shows our predictions for both cases in which the 6.4 keV K$\alpha$ emission line is included in the model or not. We can estimate reliable XZs for a number of obscured ($\log N_H/\mathrm{cm^{-2}} > 22$) AGN between $\sim$5500 and 6500, and for between 1600 and 2000 of Compton-thick AGN. Of these, around 6 Compton-thick AGN are in the redshift bin 4 to 6.5.
Furthermore, as shown by Figure \ref{fig:results}, with additional observations will be possible to determine XZs even at higher redshifts, as long as enough photons will be detected.
It is worth noting that our simulated mock datasets do not assume a redshift-dependent evolution of the $\log N_H/\mathrm{cm^{-2}} > 23$ population. However, numerous observations suggest that as we look at high redshifts, this population could make up as much as 90\% \citep[e.g.,][]{ricci17b,ananna19}, possibly due to the interstellar medium contribution to the overall absorption \citep[e.g.,][]{gilli22}. As a result, the numbers we present, especially for $z>2$, should be considered conservative lower limits.

Our results offer an alternative solution for the redshift determination of X-ray sources. Notably, this approach circumvents the reliance on the ONIR multi-band counterparts, which is becoming increasingly challenging with the advent of new data from instruments like the James Webb Space Telescope (JWST), which offers a superior angular resolution compared to X-rays. As a result, our method will substantially refine constraints on the X-ray luminosity function (XLF) and obscured AGN fractions up to redshift 6. In particular, we will be able to determine a solid Comton-thick fraction up to redshift 4 and place constraints up to z$\sim$6. This holds particular importance because current work shows discrepancies in the Compton-thick fraction even in the local Universe \citep[e.g.,][]{ricci17b, ananna19, torres-alba21} and, at high redshift ($z>4$), the use of extrapolations or XLFs characterized by notable uncertainties is common due to a lack of strong empirical support.

\section{Materials and Methods}\label{sec:4}
\begin{figure}[H]
\centering
\includegraphics[width=10cm]{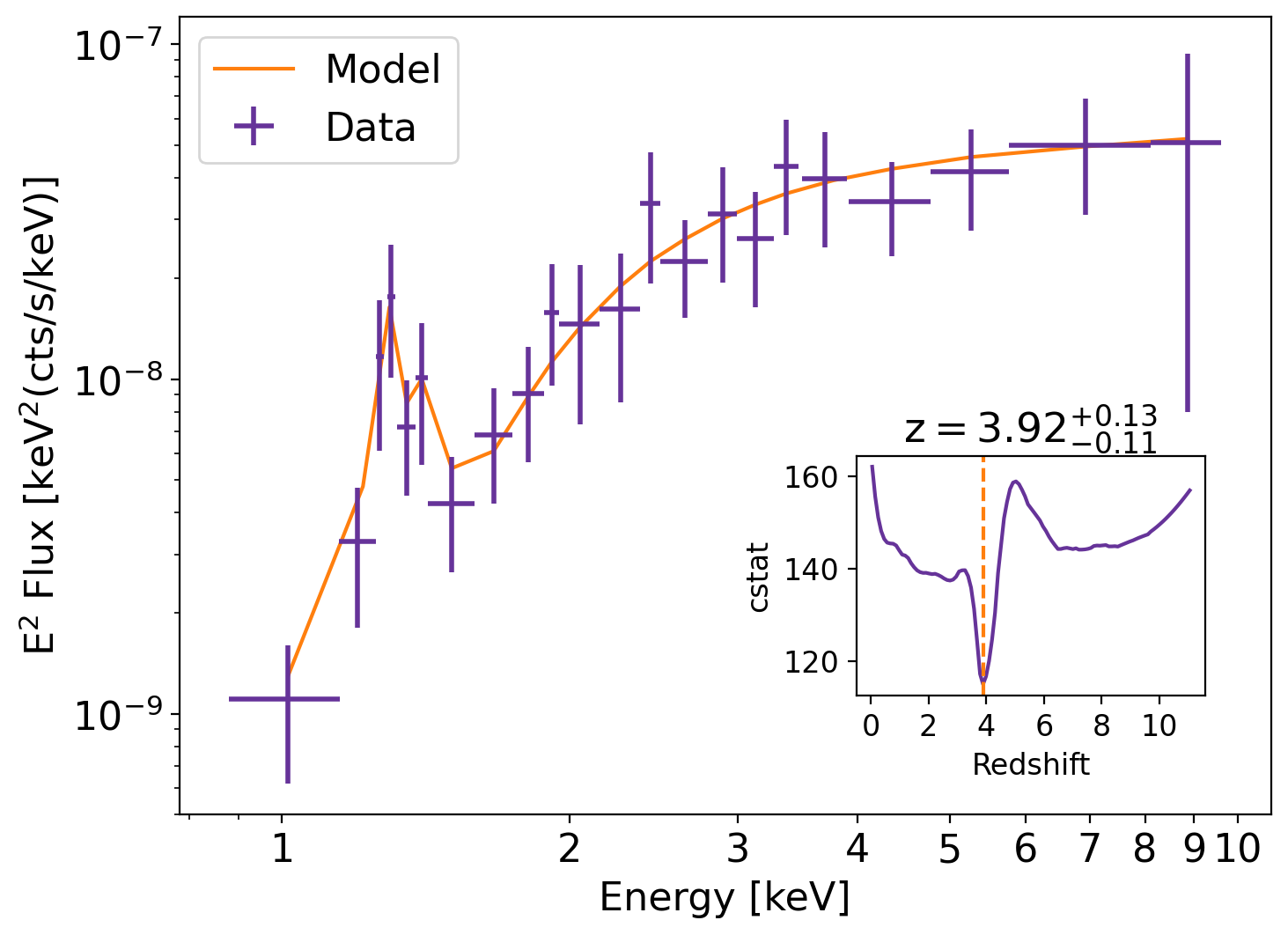}
\caption[Predicted AXIS spectrum and best-fit XZ value for a Compton-thick AGN at z=4.]{Predicted unfolded AXIS spectrum for a Compton-thick AGN at $z=4$, with $\log N_H/\mathrm{cm^{-2}} =24$, and $\log L_X/\mathrm{erg\,s^{-1}}=43.5$. The spectrum (purple points) was simulated assuming a double power-law model with the secondary power-law normalization as 3\% of the primary one (orange curve). The number of net counts is 146. The inset in the figure shows the result obtained by the \textsc{steppar} command on the redshift parameter. The minimum marks the best-fit XZ value, which agrees well with the simulated redshift.}
\label{fig:results3}
\end{figure}

The results shown in Section \ref{sec:2} were obtained using a procedure similar to \citep{peca21}. All the simulations were conducted using PyXSPEC v2.1.1 (equivalent to XSPEC v12.13.0 \citep{xspec}).
We adopted a double power-law model (\textsc{zphabs} $\times$ \textsc{zpowerlw} $+$ \textsc{zpowerlw} in XSPEC) with a fixed intrinsic photon index $\Gamma = 1.9$ and a secondary power-law normalization free to vary up to 20\% of the primary power-law normalization (as observed in X-ray surveys \citep[e.g.,][]{ricci17b,peca23}). The Fe K$\alpha$ emission line was also modeled with a redshifted Gaussian line (\textsc{zgauss}) at 6.4 keV in the rest frame, with a width of $\sigma = 10$ eV. Different line normalizations were used to obtain a canonical range of rest-frame equivalent widths, between 10 eV and 2 keV, as a function of $N_H$ \citep[e.g.,][]{ghisellini94,lanzuisi15}. An additional absorption component (\textsc{phabs}) with a fixed value of $N_H= 1.8 \times 10^{20}$ cm$^{-2}$, corresponding to the average Galactic absorption at high latitude, was added. The free parameters were the two power-law normalizations, $N_H$, line normalization, and redshift. We discuss the effects of assuming different models in Appendix \ref{app:models}.
To reproduce what is observed in intermediate-to-deep X-ray surveys \citep[e.g.,][]{liu17,marchesi16b}, we simulated column densities from $\log N_H= 22$ to $25$ with a step of 0.5, redshifts up to 9 with a step of 0.5, and different power-law normalizations to obtain a number of net counts (0.5-10 keV band) up to 150. For each parameter combination, we simulated 200 spectra. The simulations are also repeated by excluding the Fe K$\alpha$ emission line to mimic cases where no line is detected.
The expected background, which comprises both particle (non-X-ray) and astrophysical components, was properly modeled (see Appendix \ref{app:bkg} for details) and associated with the simulated spectra. The most recent AXIS responses ARF and RMF were applied. 
Simulated spectra were binned to a minimum of 1 cts/bin to avoid empty channels, and Cash statistics \citep{cash79} was applied.
Following \cite{peca21}, after a first fit to assess the free parameters, the best-fit redshift estimates (XZs) were evaluated using the \textsc{steppar} command. This allows the evaluation of the possible best-fit solutions by running consecutive fits as a function of one or more parameters. We ran it as a function of $z$ in the range [0-11], with a step of 0.01. The best-fit XZs were then determined as the minimum value in the resulting statistical distributions. Finally, the model best-fit parameters were then compared to the simulated values as described in Section \ref{sec:2}. An example of this procedure is shown in Figure \ref{fig:results3}.

\section{Conclusions}
We showed in this work how AXIS is poised to make significant contributions to the study of AGN populations, especially obscured AGN, through accurate X-ray redshift estimations. Our simulations suggest that AXIS can accurately estimate XZs for obscured AGN up to redshift $z\sim6$ in the planned deep and intermediate surveys. 
Specifically, AXIS will be able to obtain X-ray spectra with enough photons to make XZ estimates for more than 5500 obscured AGN, of which at least $\sim$1600 in the Compton-thick regime. Moreover, with additional observations it would be possible to determine XZs even at higher redshifts, as long as enough photons will be detected. 
This capability will substantially contribute to our understanding of the characteristics and evolution of heavily obscured and elusive AGN, which bear critical importance in population studies.
With the presented predictions, our approach will significantly improve the accuracy of XLFs and obscured AGN fractions up to redshift $\sim$6 and beyond. This advancement will address critical gaps in our understanding of obscured AGN both locally and at high redshifts (z$\gtrsim$4), where current discrepancies may be explained by the lack of solid observational evidence that AXIS will instead provide.

\vspace{6pt} 



\vspace{1cm}\acknowledgments{We kindly acknowledge the AXIS team for their outstanding scientific and technical work over the past year. This work is the result of several months of discussion in the AXIS-AGN SWG.}

\bigskip\medskip
\abbreviations{The following abbreviations are used in this manuscript:\\

\noindent 
\begin{tabular}{@{}ll}
AGN & Active Galactic Nuclei\\
CXB & Cosmic X-ray Background \\
MP & Match Percentage\\
NXB & Particle background \\
ONIR & Optical and Near-IR \\
SED & Spectral Energy Distribution \\
XLF & X-ray Luminosity Function \\
XZ & X-ray Redshift\\

\end{tabular}}

\appendixtitles{yes} 
\appendixsections{multiple} 
\appendix
\section{Background simulation}\label{app:bkg}
The expected background combines two main components: astrophysical and particle (NXB) backgrounds. For the latter, we used an empirical model based on the observed Suzaku XIS1 NXB spectrum \citep{tawa08}, scaled to what is expected for AXIS at low-Earth orbit. It was modeled with a series of instrumental Gaussian emission lines and three power-law components to mimic the continuum shape. Since the NXB does not depend on the effective area of the instrument, it was not convolved with the ARF during the simulations. 
The astrophysical background is the contribution of the unresolved X-ray background (CXB) and the Galactic foreground emission (local hot bubble and Milky Way hot halo). To model its contribution, we used the models provided by \citep{cappelluti07} and \citep{bluem22}, respectively.

\section{Model complexity}\label{app:models}
An important factor to consider when dealing with low photon statistics is the complexity of the model used to determine XZs. In general, with low-quality spectra only simple models are used \citep[e.g.,][]{peca21, signorini23} as it becomes too challenging to fit the larger number of components present in complex models. In addition, fixing many parameters to the default values makes the shape of complex models similar to simple ones, leading to consistent results but requiring more computational time \citep[e.g.,][]{peca23}.
In this Section \ref{sec:4}, we show the results obtained with a double power-law model. \citep{peca21} show how using a more simple, single power-law model, and a more complex model such as MYTorus \citep{murphy09}, change the results on MP by no more than $\pm$15\%. Given that, we can conclude that assuming simple models for the redshift estimate of obscured AGN is a reasonable choice. The main spectral features, such as the Fe K$\alpha$ emission line at 6.4 keV and the main absorption features, are those that drive the XZ estimates and are already included in these models. Therefore, it is not necessary to add more complexity to the models to derive XZs.

\subsection{Iron K{\texorpdfstring{$\alpha$}{}} emission line and absorption features}
XZ estimates are driven by a combination of emission (notably the Fe 6.4 keV K$\alpha$ emission line) and absorption features (primarily the 7.1 Fe K$\alpha$ absorption edge and the photoelectric cut-off). However, in situations of limited photon statistics, the detection of the Fe 6.4 keV K$\alpha$ emission line is not always successful \citep[e.g.,][]{marchesi16b,signorini23}. To account for such scenarios, we run again the simulations without the Gaussian component. On average, this modification led to a decrease of approximately $\sim$10-15\% in the MP. This outcome can be explained by the fact that the larger uncertainties derived from absorption features only can cause some XZs to not meet the criterion of $|\Delta z| > 0.15(1 + z_{sim})$. In fact, although absorption is the primary determinant for XZ \citep{simmonds18,peca21}, the narrower profile of the Fe 6.4 keV K$\alpha$ feature allows for smaller errors \citep{peca21,signorini23}. During the computation of uncertainties, XSPEC evaluates the statistics around the best-fit solution, and, consequently, when an emission line is identified, the X-ray redshift likelihood experiences a rapid decline before and after the best-fit value, resulting in smaller uncertainties. In contrast, when only broad absorption features are identified, the uncertainties on XZs are larger.


\externalbibliography{yes}
\bibliography{template}

\end{document}